\documentclass{ws-p9-75x6-50}

\begin{document}

\title{Bulge formation in very late-type galaxies}

\author{YanNing Fu}

\address{Purple Mountain Observatory, Chinese Academy of Sciences, Nanjing, China}

\author{JieHao Huang}

\address{Department of Astronomy, Nanjing University, Nanjing, China}

\author{ZuGan Deng}

\address{Department of Physics, Graduate School, Chinese Academy of Sciences, Beijing, China}


\maketitle

\abstracts{
The dynamical evolution of super star clusters 
(SSCs) moving in the background of a dark matter halo has been investigated 
as a possible event responsible for the formation of bulges in 
late-type spirals. The underlying physical processes include  sinking
of SSCs due to the dynamical friction and  stripping of SSCs on their 
way  to the center.  Based on the assumption of a universal density profile 
for the dark matter halo, and an isothermal model for the SSCs, our 
simulations have yielded bulges that are similar in many aspects to 
the observational ones. In particular, the derived surface density profiles 
can be well fitted by an exponential structure with nuclear cusps,
which is consistent with HST observations.The preliminary simulations with 
the Burkert density profile yield very instructive predictions on 
the bulge formation, which is surely worth further investigating.}

\section{Introduction}
It is believed for a time that bulges in many late-type galaxies
display exponential surface brightness profiles. Now the HST observations
(Carollo \& Stiavelli 1998; Carollo 1999) revealed a very interesting 
central  structure in a samll sample of late-type galaxies, which
illustrates typical surface brightness profiles of nucelar cusps on 
top of exponential bulges. Carollo and her colleagues convincingly claimed 
the stellar origin for these nuclear cusps.

In fact, nuclear star clusters may be a common phenomenon 
(e.g. Davidge \& Courteau 2002;  and references therein)  in late-type
galaxies. Possibly related to the existence of nuclear star
clusters is the presence of circumnuclear young massive star
clusters, or circumnuclear SSCs.

A question then arises whether there is a physical
relation between these two common phenomena in late-type spirals,
i.e., the common existence of nuclear/circumnuclear star clusters
and the exponential bulges with nuclear stellar cusps. If this is
the case, what is the physical process responsible for this
relation? A prevailing view for the formation of bulges in
late-type galaxies is the secular evolution of disks due to the
bar formation/destruction (e.g. Norman, Sellwood, \& Hasan 1996).
However, Bureau argued in his review (2002) against this scenario
based on the consideration of the fast duty cycle of the bar
formation/destruction, inferred from the omnipresence of bars.

In this paper we report our study, motivated by Carollo's claims
and the observations mentioned above, on the dynamical evolution
of circumnuclear SSCs moving in the dark matter halo as a possible
process for the formation of bulges with nuclear cusps in very
late-type galaxies.

\section{Models}

\subsection{SSC model}

Based on the analogy with globular clusters (GCs), we assume that SSCs 
have a similar mass spectrum as the initial GC mass function but with 
larger mean value. We adopt the 
log-normal mass function as SSCs' mass function (SSCMF) 
\begin{equation}
\label{SSCmassspectrum} Log_{10}(\frac{M}{M_{\odot}}) \sim  N(
\exp(mean)=2 \times 10^6 M_{\odot}, variance=0.08)
\end{equation}
Considering the mean GC's mass of about $1-2
\times 10^5 M_{\odot}$, we may infer the mean SSC's mass of about
$2 \times 10^6 M_{\odot}$.

\subsection{Background}
Considering the
bulge-formation process in very late-type galaxies (Sd type and
later) we only take a dark matter (DM) halo as the initial background. For
the dark halo, we assume the universal density profile (Navarro et
al. 1997, hereafter NFW), which can be written as (e.g., Binney \& Merrifield,
1998)
\begin{equation}
\label{NFWMassdensity} \rho_h(r)=\frac{M_{0h}}{r(a_h+r)^2}
\end{equation}
where $r$ is the distance from the halo center.

We consider two halos with quite
different masses:  a large one with mass
$M_{200}=10^{12} h^{-1} M_{\odot}$
 and a smaller halo with mass
$M_{200}=3 \times 10^{11} M_{\odot}$. 
The NFW halos are  determined via the scaling law  
(see Table 1), 
with the Hubble constant $H_{0} = 75~{\rm km s^{-1} Mpc^{-1}}$, 
$h$ = 0.75.
\vspace{3mm}
\begin{center}
{\bf Table 1.} Parameters of the chosen NFW halos
\end{center}
\vspace{1mm}
\begin{center}
\begin{tabular}{|c|c|c|c|c|}
\hline
$M_{200} (M_{\odot})$ & $r_{200} (kpc)$ & $c$ & $a_{h}(kpc)$ & $M_{0h}(M_{\odot})$ \\
\hline
$10^{12}h^{-1}$ & 216.8 & 8.5 & 25.5 & $7.8 \times 10^{10}$ \\
\hline
$3 \times 10^{11}$ & 131.9 & 10 & 13.2 & $1.6 \times 10^{10}$ \\
\hline

\end{tabular}
\end{center}
\vspace{3mm}

In order to account for the background variation caused 
by the fallen SSCs together with their
stripped mass,
we add a spherical component, represented by shells of equal
density, which varies according to the changing SSC mass
contribution.

\subsection{Dynamical friction}
We define  $M$ and $\vec{V}_M$ (with $V_M=|\vec{V}_M|$)
equal to, respectively, the mass and velocity of cluster
experiencing the dynamical friction. Assuming the background
matter has a Maxwellian
velocity distribution with dispersion $\sigma_{bkgd}$, and is
composed of particles with mass much smaller
than $M$, the dynamical friction formula may be written as
 (e.g., Binney \& Tremaine, 1987)
\begin{equation}
\label{Chdf} \frac{d\vec{V}_M}{dt}=-\frac{2 \pi \log(1+\Lambda^2)
G^2 M \rho}{V_M^3}[ {\rm erf} (X)-\frac{2X}{\sqrt{\pi}}
\exp(-X^2)] \vec{V}_M
\end{equation}
where {\rm erf} is the error function. The readers are advised on
 referring to Binney \& Tremaine (1987) about  $\Lambda~ and~ X $.

\subsection{Stripping}
We assume that the stellar mass outside a sphere of radius
$R_t$, which corresponds to the instantaneous Hill stability
region around the SSC center, will be stripped.
As a
first-order approximation, the mass of the stripped stars is
considered to be, at some later epoch, uniformly distributed in
the shell bounded by $r+R_t(r)$ and $r-dr-R_t(r-dr)$. By summing
up all of the stellar mass stripped at various $r$s, the stripped
stellar mass distribution can be derived.

If a single massive object is embedded in the center of an
SSC, stripping cannot proceeded further when only this object
is left. Therefore, in
our simulations, we consider two extreme cases:  stripping is
not allowed when the mass of the stripped SSC is less than $1
M_{\odot}$ and $1000M_{\odot}$, respectively.

\section{Results}

Figure 1 shows the mean  surface density profiles of simulated 
galaxies with high-density central components formed at 1 Gyr (with a 
proportion of about 65\% and 90\% for the larger and smaller
halos, respectively).

\begin{figure}
\includegraphics[width=8cm]{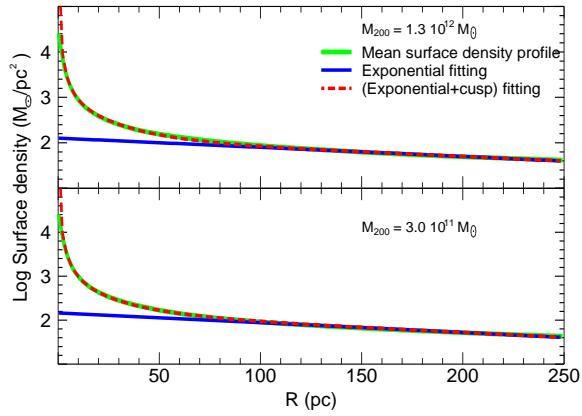}
\caption{Mean surface densities derived by SSCs without IMBHs for two DM density profiles adopted. The model fittings are illustrated with solid blue and dashed red lines}
\end{figure}

In order to give a quantitative comparison with the HST
observations, the predicted mean surface density profiles are fit with 
a model, following Carollo \& Stiavelli (1998), which is given by

\begin{equation}
\label{model} \sigma(R)=\sigma_0 \exp(-1.678
\frac{R}{R_e})+\sigma_1(1+\frac{R_c}{R})^\gamma
\exp(-\frac{R}{R_s})
\end{equation}

The resulting profiles are also shown in Fig. 1, and the associated
values of the parameters are given in Table 2.
Except at radii less than about 2 pc, the fittings shown in Fig. 1
are quite satisfactory. The half mass radii of the formed bulges are 
about 300pc, and the scale-sizes of the nuclear cusps are about 10pc ,
which  are consistent with HST observations.
\vspace{3mm}
\begin{center}
{\bf Table 2.} The fitted values of parameters in (\ref{model})
\end{center}
\vspace{1mm}
\begin{center}
\begin{tabular}{|c|c|c|c|c|c|c|}
\hline $M_{200}$~$(M_{\odot})$  & $\sigma_0$~($M_{\odot}/pc^2$) & $R_e$~($pc$) & $\sigma_1$~($M_{\odot}/pc^2$) & $R_c$~($pc$) & $\gamma$ & $R_s$~($pc$) \\
       
\hline $10^{12} h^{-1}$ & 127 & 361 & 317 & 8 & 2.2 & 23 \\
 $3 \times 10^{11}$ & 146 & 327 & 212 & 18 & 1.7 & 27 \\
\hline
\end{tabular}
\end{center}
\vspace{3mm}
In order to see what could happen to the galaxies
where no high-density central component is formed at 1 Gyr, we
extend the simulation to 1.5 Gyr. The bulge formation rates are
then increased to about 90\% and 99\% for the larger and smaller
halos, respectively. This result implies that the proposed
mechanism  may be a viable mode  
of bulge formation.

\section{Discussion}

Now more and more evidence seems to argue  (e.g. Marchesini et al.  2002
and the references therein) that
the NFW profiles are not adequate for a  fraction of dwarf galaxies  
 which are  dominated  by the dark matter halos, and
suggested  that the Burkert profile is more suitable to these  galaxies.
We  would then take the Burkert density profile given below for our simulations.

\begin{equation}
\label{BurkertMassdensity} \rho_{DM}(r)=\frac{\rho_o~r^{3}_{o}}{(r+r_o)(r^{2}+r^{2}_{o})}
\end{equation}

The work of this kind has not been finished yet.  One of
our preliminary results is illustrated in Fig 2, obtained 
with a  central density of 0.10 $M_\odot~pc^{-3}$ .
It is very instructive to notice that 
the exponential bulges with nuclear cusps formed in about 40\%  galaxies,
and a large fraction of dwarf spirals  have no bulges and no nuclear cusps 
formed.

\begin{figure}
\includegraphics[width=8cm]{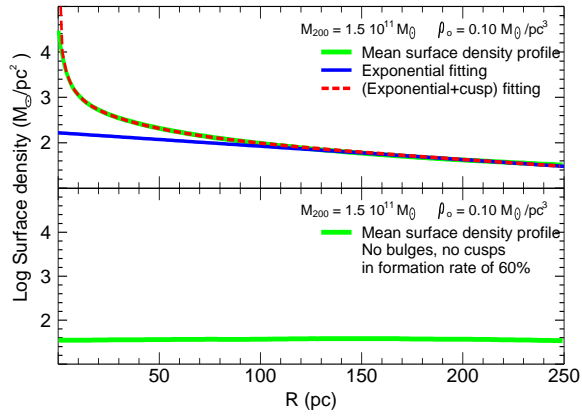}
\caption{Model fitting to the simulations with the Burkert density profiles }
\end{figure}

\vspace{3mm}
\begin{center}
{\bf Table 3.} Comparisons of the predicted to the observed bulge formation rates
\end{center}
\vspace{1mm}
\begin{center}
\begin{tabular}{|c|c|c|c|}
\hline
 & NFW & Burkert & Obs.($\sim$Sd)$^{*}$\\
 &     &0.03\hspace{3mm}($\rho_{o}$)\hspace{3mm}0.10&\\
\hline
bulge & $\sim$90\%&0\%\hspace{10mm}$\sim$40\%& $<$10\%~$-$~50\% \\
 size& 300~pc &\hspace{18mm}200pc & N/A\\
 cusp & $\sim$90\% & 0\%\hspace{10mm}$\sim$40\%&$\sim$30\%~$-$~50\%\\
 size&8$-$10~pc&\hspace{18mm}12pc& $<$1pc~$-$ $<$20pc\\
\hline

\end{tabular}
\vspace{2mm} \\
$^{*}$  based on Matthews et al. (2002, 1999) and   Carollo (1999)\\
\end{center}
\vspace{3mm}

In contrast to the case of  NFW density profiles, the bulge formation rates
yielded with the Burkert ones are obviously lower.
To compare both of these predicted rates to
the relevant data obtained for very late-type galaxies might
give a clue to analysing the DM density profiles adequate 
to these  galaxies. The comparisons summarized in  Table 3
apparantly indicate that the Burkert density profile is more suitable
to the galaxies observed by Matthews et al. (2002, 1999) and   Carollo (1999).

Nevertheless, we have to be cautious of the above claim. For one thing, 
the observed properties for galaxies with Hubble type of Sd or later
are taken from a small sample of about 20 sources only. And more important 
is that the results obtained for the Burkert profiles are very  preliminary.
A detailed investigation on this topic is underway.

\end{document}